\documentclass[conference]{IEEEtran}

\ifCLASSINFOpdf
\else
\fi
\usepackage{amsmath}
\usepackage{amsthm}
\usepackage{amssymb}
\newcommand{\be}{\begin{equation}}
\newcommand{\ee}{\end{equation}}
\usepackage{graphicx}
\begin{document}
%
\title{Fractional Dynamical Systems}

\author{\IEEEauthorblockN{Mark Edelman}
\IEEEauthorblockA{ Department of Physics\\
Stern College at Yeshiva University\\
 245 Lexington Avenue, New York, NY 10016, \underline{USA} and  \\
Courant Institute at New York University \\
        251 Mercer Street, New York, NY 10012, \underline{USA}
}
Email: edelman@cims.nyu.edu}


%


\maketitle

\begin{abstract}
In this paper the author presents the results of the preliminary investigation of fractional dynamical systems based on the results of numerical simulations of fractional maps.  Fractional maps are equivalent to fractional differential equations describing systems experiencing periodic kicks. Their properties depend on the value of two parameters: the non-linearity parameter, which arises from the corresponding regular dynamical systems; and the memory parameter which is the order of the fractional derivative in the corresponding non-linear fractional differential equations. 
The examples of the fractional Standard and Logistic maps demonstrate that phase space of non-linear fractional dynamical systems may contain periodic sinks, attracting slow diverging trajectories, attracting accelerator mode trajectories, chaotic attractors, and cascade of bifurcations type trajectories whose properties are different from properties of attractors in regular dynamical systems. 
The author argues that discovered properties should be evident in the natural (biological, psychological, physical, etc.) and engineering systems with power-law memory. 
\end{abstract}

\IEEEpeerreviewmaketitle

\section{Introduction}
\label{I}

Fractional differential equations are used in applications to describe systems with long-range interactions \cite{LRBook,LR1,LR2} or systems with power-law memory. Systems with memory include: Hamiltonian systems where memory is the result of stickiness of trajectories in time to the islands of regular motion \cite{ZBook,ZEN}; dielectric materials where electromagnetic fields  are described by equations with time fractional derivatives due to the ’universal’ response - the power-law frequency dependence of the dielectric susceptibility in a wide range of frequencies \cite{TBook,TD1,TD2,TD3}; viscoelastic materials and materials with rheological properties where non-integer order differential stress-strain relations give a minimal parameter set concise description of polymers and other viscoelastic materials with non-Debye relaxation and memory of strain history 
\cite{MBook,VR1,VR2,VR3,VR4}. Examples of fractional order systems in modeling and control may be found in \cite{CB}.

The intuitively obvious fact that many biological systems are systems with memory is now confirmed by rigorous research. In many cases memory obeys the power law and the corresponding systems could be described by fractional differential equations.
It has been shown recently that viscoelastic properties are typical for a wide variety of biological tissues (liver, heart valve, brain, etc.) \cite{BV1,BV2,BV3,BV4,BV5,BV6} which implies that they can be described by differential equations with time fractional derivatives. 
Short- and long-term memory effects in neuronal circuits may result from synaptic plasticity \cite{Neuron1} and from intrinsic membrane currents 
\cite{Neuron2}. As it has been shown recently, even processing of external stimuli by individual neurons can be described by fractional differentiation \cite{Neuron3,Neuron4}.
In some recent papers describing dynamics of biological systems authors used power-law adaptation \cite{Ada1,Ada2,Ada3,Ada4,Ada5}.
Fractional differentiation is used in bioengineering research (bioelectrodes, biomechanics, bioimaging) \cite{Magin}.
The power-law memory $t^{-\beta}$ with $\beta \in (0,1)$ was demonstrated in experiments on human learning and forgetting 
\cite{Anderson,Kahana,F1,F2,F3,F4}.

Fractional differential equations are integro-differential equations.
Their numerical solution requires large computer memory and long runs of numerical simulations which makes it very difficult to investigate general properties of fractional dynamical systems - systems which can be described by fractional differential equations.  
Investigation of general properties of nonlinear fractional dynamical systems in \cite{FM1,FM2,FM3,FM4,FM5,FM6,FM7} has been based on fractional maps derived from fractional differential equations describing systems experiencing periodic kicks introduced in \cite{FM1,FM2,FM5,FM8,FM9}. 
In the following sections we'll review fractional maps
(Sec. \ref{FM}) and some of their properties (Sec. \ref{FA}).
In Conclusion we'll discuss possible manifestations of fractional dynamical systems in nature.

\section{Fractional Maps}
\label{FM}

The notion of the $\alpha$-Families of Maps ($\alpha$-FM) was introduced in \cite{FM5}.

\subsection{Universal Fractional Map}
\label{UFM}

The derivation of the equations of the Universal $\alpha$FM starts with the equation:  
\begin{equation}
\frac{d^{\alpha}x}{dt^{\alpha}}+G_K(x(t- \Delta)) \sum^{\infty}_{k=-\infty} \delta \Bigl(t-(k+\varepsilon)
\Bigr)=0,   
\label{UM1D2Ddif}
\end{equation}                                                       where $\varepsilon > \Delta > 0$,  $\alpha \in \mathbb{R}$, $\alpha>0$, 
$\varepsilon  \rightarrow 0$, and the initial conditions corresponding to the type of fractional derivative to be used. 
 
In the case of the Riemann-Liouville fractional derivative integration of Eq.~(\ref{UM1D2Ddif}) with the initial conditions
\be
(_0D^{\alpha-k}_tx)(0+)=c_k, 
\label{ic}
\ee                                                                   where $k=1,...,N$, $ N-1 < \alpha \le N$, and $N \in \mathbb{N}$, produces the Riemann-Liouville Universal $\alpha$-FM
{\setlength\arraycolsep{0.5pt}
\begin{eqnarray}
&&p^s_{n+1}= p^s_n + \sum^{N-s-3}_{k=0}\frac{p^{k+s+1}_n}{(k+1)!} 
-\frac{G_K(x_n)}{(N-s-2)!}, 
\label{FrRLMappConv} \\
&&x_{n+1}=  \sum^{N-1}_{k=2}\frac{c_k}{\Gamma(\alpha-k+1)}(n+1)^{\alpha -k} \nonumber \\  
&&+\frac{1}{\Gamma(\alpha)}p^{N-2}_{n+1}+ \frac{1}{\Gamma(\alpha)} \sum^{n-1}_{k=0} p^{N-2}_{k+1}V^1_{\alpha}(n-k+1), 
\label{FrRLMapxConv} 
\end{eqnarray} 
}
where $s=0,1,...N-2$ and $V^k_{\alpha}(m)=m^{\alpha -k}-(m-1)^{\alpha -k}$.
In Eqs.~(\ref{FrRLMappConv})~and~(\ref{FrRLMapxConv}) $p(t)= {_0D^{\alpha-N+1}_t}x(t)$, 
$p^{(s)}(t)= {D^{s}_t}p(t)$, $x_{n}=x(n)$, and $p^{(s)}_{n}=p^{(s)}(n)$.

In the case of the Caputo fractional derivative integration of Eq.~(\ref{UM1D2Ddif}) with the initial conditions
\be
(D^{k}_tx)(0+)=b_k,
\label{cic}
\ee                                                                   where $k=0,...,N-1$ produces the Caputo Universal $\alpha$-FM

{\setlength\arraycolsep{0.5pt}
\begin{eqnarray}
&&x^{(s)}_{n+1}= \sum^{N-s-1}_{k=0}\frac{x^{(k+s)}_0}{k!}(n+1)^{k} \nonumber \\ 
&&-\frac{1}{\Gamma(\alpha-s)}\sum^{n}_{k=0} G_K(x_k) (n-k+1)^{\alpha-s-1},
\label{FrCMapx}
\end{eqnarray} 
}
where $s=0,1,...,N-1$ and $x^{(s)}(t)=D^s_tx(t)$.

For integer $\alpha = N$ the Universal $\alpha$-FM can be written in the form of the N-dimensional volume preserving (for $N \ge
2$) map
{\setlength\arraycolsep{0.5pt}
\begin{eqnarray}
&&p^s_{n+1}= p^s_n + \sum^{N-s-3}_{k=0}\frac{ p^{k+s+1}_n}{(k+1)!} 
-\frac{G_K(x_n)}{(N-s-2)!}; 
\label{IntRLMappConv} \\
&&x_{n+1}=   x_n + \sum^{N-2}_{k=0}\frac{p^{k}_n}{(k+1)!} 
-\frac{G_K(x_n)}{(N-1)!}.
\label{IntRLMapxConv} 
\end{eqnarray} 
}
For $N=\alpha=2$ and $G_K(x)=KG(x)$ this map gives the well-known in regular dynamics 2D Universal Map \cite{ZBook}
\begin{equation}
p_{n+1}= p_{n} - KG(x_n),
\label{UMp}
\end{equation}
\begin{equation}
x_{n+1}= x_{n}+ p_{n+1}.
\label{UMx}
\end{equation}

\subsection{Standard and Logistic Fractional Maps}
\label{SLFM}

In the case $G(x)= \sin(x)$ Eqs.~(\ref{UMp})~and~(\ref{UMx}) produce
the well known Standard Map \cite{Chirikov}
\begin{equation}
p_{n+1}= p_{n} - K \sin(x_n), \ \ \ ({\rm mod} \ 2\pi ), 
\label{SMp}
\end{equation}
\begin{equation}
x_{n+1}= x_{n}+ p_{n+1}, \ \ \ ({\rm mod} \ 2\pi ).
\label{SMx}
\end{equation}
This is why the Universal $\alpha$-Families of Maps with
\begin{equation}
G_K(x)=K \sin(x)
\label{SFM}
\end{equation}                                                         are called the Riemann-Liouville Standard $\alpha$-FM 
(Eqs.~(\ref{FrRLMappConv})~and~(\ref{FrRLMapxConv})) and the Caputo Standard $\alpha$-FM (Eq.~(\ref{FrCMapx})) correspondingly.

For $\alpha=1$ the Caputo Universal $\alpha$-FM Eq.~(\ref{FrCMapx})
\begin{equation}
x_{n+1}=x_n-G_K(x_n)
\label{1DUM}
\end{equation}                                                      produces the regular Logistic Map $x_{n+1}=Kx(1-x)$ if 
\begin{equation}
G_K(x)=G_{LK}(x)=x-Kx(1-x).
\label{LFM}
\end{equation}
This is why the Univeral $\alpha$-FMs with $G_K(x)=G_{LK}(x)$ are called the Logistic $\alpha$-FMs.

The Standard and Logistic $\alpha$-FMs were numerically investigated for $0<\alpha \le2$, the case which is most important for applications.  
\begin{itemize}
\item{
For $0<\alpha <1$ the Caputo Standard $\alpha$-FM can be written as
\begin{equation}
x_{n}=  x_0- 
\frac{K}{\Gamma(\alpha)}\sum^{n-1}_{k=0} \frac{\sin{x_k}}{(n-k)^{1-\alpha}},
 \   \  ({\rm mod} \ 2\pi ).
\label{FrCMapSM}
\end{equation}}
\item{
The Caputo Logistic $\alpha$-FM for $0<\alpha <1$ assumes the form
\begin{equation}
x_{n}=  x_0+ 
\frac{1}{\Gamma(\alpha)}\sum^{n-1}_{k=0} \frac{Kx_k(1-x_k)-x_k}{(n-k)^{1-\alpha}}.
\label{FrCMapLM}
\end{equation}
}
\item{
The 1D Standard Map can be written as the Circle Map with zero driving phase
\begin{equation}
x_{n+1}= x_n - K \sin (x_n), \ \ \ \ ({\rm mod} \ 2\pi ). 
\label{SM1D} 
\end{equation}
}
\item{
For $1<\alpha <2$ the Riemann-Liouville Standard $\alpha$-FM can be written as
\be
p_{n+1} = p_n - K \sin x_n ,  \label{FSMRLp}
\ee
\be
x_{n+1} = 
\frac{1}{\Gamma (\alpha )} 
\sum_{i=0}^{n} p_{i+1}V^1_{\alpha}(n-i+1) 
,\ \ ({\rm mod} \ 2\pi ). \label{FSMRLx} 
\ee
}
\item{
The Caputo Standard $\alpha$-FM for $1<\alpha <2$ assumes the form
{\setlength\arraycolsep{0.5pt}   
\begin{eqnarray}
&& p_{n+1} = p_n - 
 \frac{K}{\Gamma (\alpha -1 )} 
\Bigl[ \sum_{i=0}^{n-1} V^2_{\alpha}(n-i+1) \sin x_i + \nonumber \\
&& \sin x_n \Bigr],\ \ ({\rm mod} \ 2\pi ),  \label{FSMCp}  \\ 
&& x_{n+1}=x_n+p_0-
\frac{K}{\Gamma (\alpha)} 
\sum_{i=0}^{n} V^1_{\alpha}(n-i+1) \sin x_i, \nonumber \\ 
&& ({\rm mod} \ 2\pi ). 
\label{FSMCx}
\end{eqnarray}
}}
\item{
For $1<\alpha <2$ the Riemann-Liouville Logistic $\alpha$-FM can be written as
{\setlength\arraycolsep{0.5pt}
\begin{eqnarray}
&&p_{n+1} = p_n - Kx_n (1-x_n)-x_n,  \label{LMRLp}  \\
&&x_{n+1} = \frac{1}{\Gamma (\alpha )} 
\sum_{i=0}^{n} p_{i+1}V^1_{\alpha}(n-i+1). \label{LMRLx} 
\end{eqnarray} 
}}
\item{
The Caputo Logistic $\alpha$-FM for $1<\alpha <2$ assumes the form
{\setlength\arraycolsep{0.5pt}
\begin{eqnarray}
&&x_{n+1}=x_0+ p(n+1)^{k} - \frac{1}{\Gamma(\alpha)}\sum^{n}_{k=0} [x_k-
\nonumber \\
&&Kx_k(1-x_k)] (n-k+1)^{\alpha-1}, 
\label{LMCx}  \\
&&p_{n+1}=p_0 - \frac{1}{\Gamma(\alpha-1)}\sum^{n}_{k=0} [x_k- \nonumber \\
&&Kx_k(1-x_k)] (n-k+1)^{\alpha-2}.
\label{LMCp}
\end{eqnarray} }}
\end{itemize} 
In the following section 
(Sec.\ref{FA}) we'll present a brief summary of the results of investigation of the  Standard and the Logistic  $\alpha$-FMs for $0<\alpha <2$ obtained in \cite{FM1,FM2,FM3,FM4,FM5,FM6,FM7}.

\section{Fractional Attractors}
\label{FA}

\begin{figure}[!t]
\centering
\includegraphics[width=3in]{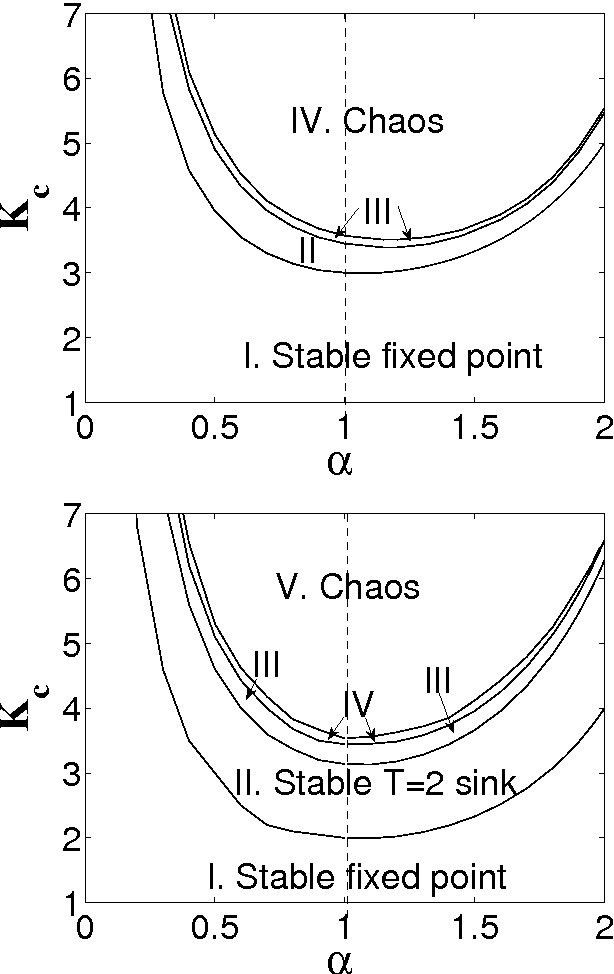}
\caption{$\alpha-K$ diagrams for the Logistic (upper figure) and the Standard (lower figure) $\alpha$-FMs. The $(0,0)$ fixed point is stable in the areas I. The $T=2$ sink is stable in the areas II.  In the area III in the bottom figure the $T=4$ sink is stable. The area III in the  upper figure is the area of the stable $T=2^n$ ($n>1$, $n \in \mathbb{N}$) sinks and cascade of bifurcations type trajectories (CBTT).  The area IV in the lower figure is the area of the 
stable $T=2^n$ ($n>2$, $n \in \mathbb{N}$) sinks and CBTT. The area IV in the upper figure and the area V in the lower figure is the area of chaos.
  }
\label{Cr}
\end{figure}

Our analysis of fractional maps for the most important in applications case $0<\alpha<2$ will be based on Fig.~\ref{Cr}
\begin{figure}[!t]
\centering
\includegraphics[width=3in]{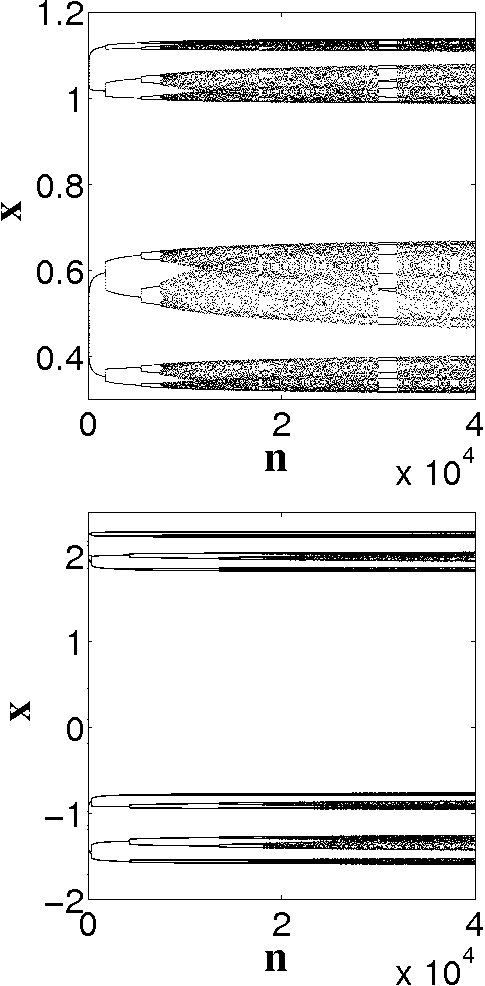}
\caption{ Examples of CBTT for 
$0<\alpha<1$ ($\alpha=0.2$): the Logistic Map with $K=11.4$ (upper figure) and the Standard Map with $K=13.05$ (lower figure). 
}
\label{CBTT1D}
\end{figure}
For small values of the non-linearity parameter $K$ the fixed point $(0,0)$ (which is a sink) is stable (area I). In the 2D phase space of the fractional Standard Map for $1<\alpha<2$ the higher period sinks which can be traced to the corresponding islands of stability of the regular Standard Map are also present. 
With the increase in $K$ the fixed point becomes unstable and the stable $T=2$ point appears (area II). The further increase in $K$ leads to the cascade of bifurcations, when the period $2^n$ ($n>1$) sinks become unstable and the stable period $2^{n+1}$ sinks appear (area III on the upper figure and III and IV on the lower). Properties of fractional sinks and their differences from properties of sinks in regular dissipative dynamical systems are discussed in the above-mentioned papers. 

Convergence of trajectories to fractional sinks obeys the power law. The exponents of this power law, defined by the memory parameter $\alpha$, may be different for trajectories with initial conditions from basins of attraction (fast convergence) and those that start in chaotic areas (slow convergence). In addition, phase space of fractional maps may contain attracting slow diverging trajectories (power-law divergence with the exponents depending on $\alpha$), attracting accelerator mode trajectories, and chaotic attractors.

Just below the border with chaos (the upper curves in Fig.~\ref{Cr}) all trajectories converge to cascade of bifurcations type trajectories (CBTT). 
\begin{figure}[!t]
\centering
\includegraphics[width=3in]{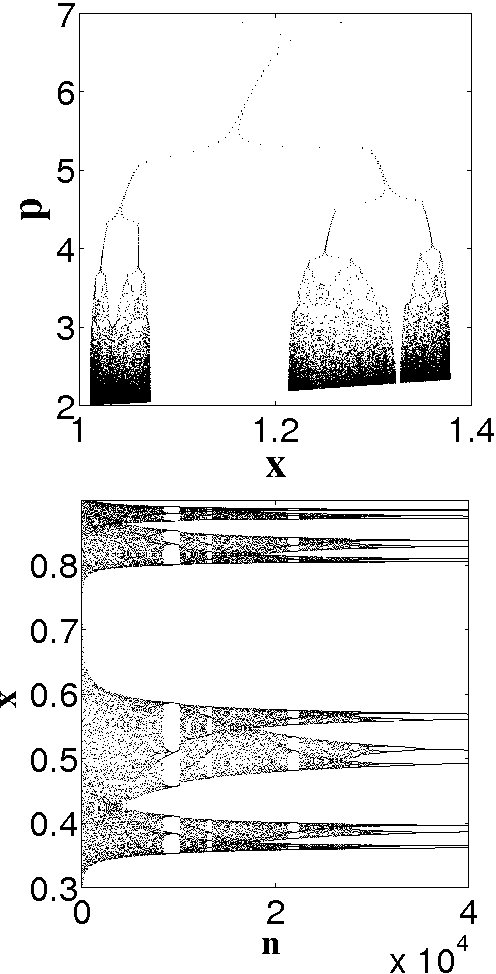}
\caption{ Examples of CBTT for 
$1<\alpha<2$: the Standard Map with $\alpha=1.3$ and $K=3.62$ (upper figure) and the inverse CBTT for Logistic Map with $\alpha=1.15$ and $K=3.45$ (lower figure). }
\label{CBTT2D}
\end{figure}
The examples of CBTT are presented in Figs.~\ref{CBTT1D}~and~\ref{CBTT2D}. In CBTT after a small number of iterations a trajectory converges to a period one trajectory (fixed point) which later bifurcates and becomes a $T=2$ trajectory and then follows the period doubling scenario typical for cascades of bifurcations in regular dynamics. The difference is that in regular dynamics a cascade of bifurcations is the result of a change in the non-linearity parameter and in CBTT a cascade of bifurcations occurs on a single attracting trajectory.
$T>1$ trajectories may also follow the cascade of bifurcations scenario which leads to a stable high-period sink. In the fractional Logistic Map with $1<\alpha<2$ instead of CBTT only inverse CBTT (lower figure in Fig.~\ref{CBTT2D}) were found.

\begin{figure}[!t]
\centering
\includegraphics[width=3in]{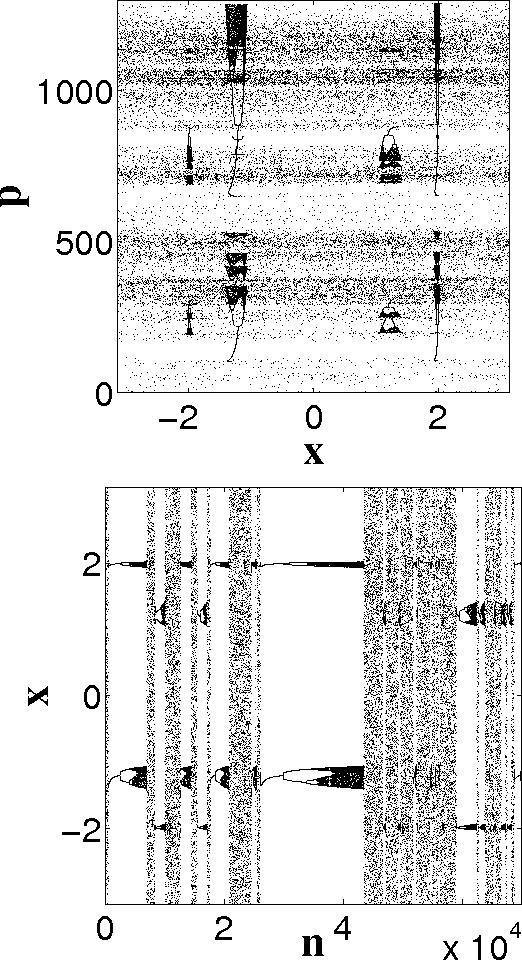}
\caption{ An example of intermittent CBTT. Standard Map with $\alpha=1.557$ and $K=4.21$.}
\label{Intermittent}
\end{figure}

An interesting phenomenon is intermittent CBTT (Fig.~\ref{Intermittent}). 
In this case a trajectory demonstrates intermittent CBTT and chaotic behavior. The time spent in CBTT may vary (see lower figure in Fig.~\ref{Intermittent}). This behavior is similar to the behavior of chaotic trajectories of regular dynamics in the presence of sticky islands. The difference is that in the latter case the trajectories spend some time in the neighborhood of the islands of regular motion and in the former case there are no stable regular trajectories.

The analysis presented in this section is far from being exhaustive, but it is enough to make some conclusions presented in the following Sec.~\ref{C}. More detailed analysis of the properties of fractional maps can be found in \cite{FM1,FM2,FM3,FM4,FM5,FM6,FM7}.

\section{Conclusion}
\label{C}

The thing that interests the author of this paper the most is how general properties of systems with power-law memory (non-uniqueness of solutions, cascade of bifurcations type trajectories, etc.) reveal themselves in nature, especially in biological systems which are almost always systems with memory (see Sec.~\ref{I}). 
Natural systems are very complicated with many complex processes going on all the time, but looking at a system as a whole     (a neuron, the brain, the heart, a human being, a society, ...) one may consider it as a system with memory (in many cases power-law) and many periodic processes. In many circumstances one may separate the most important process and then a fractional differential equation with periodic kicks could be an appropriate model for the considered system.

Certainly, there is a lot of noise and randomness in a natural system. 
How will it modify the system's behavior? This is one of the problems to solve.
The conjecture is that up to a certain level of noise trajectories would wander in the proximity of the unperturbed solutions and cascade of bifurcations type trajectories would turn into sets of bands in which a system will evolve.

Most of the beautiful features one could expect considering biological systems with memory, including cascade of bifurcations type behavior, would exist only if the system is sufficiently non-linear. Non-linearity is characterized by the value of a certain parameter. If the value of this parameter is small, the system is stable. When the value reaches a critical point a cascade of bifurcations type behavior appears: system first exists in the vicinity of a stable fixed point; after some time (which depends on the value of the non-linearity parameter and memory) it bifurcates and may stay in a new period two state forever or may bifurcate again and so on. The process may end in chaos. 
For certain values of the system parameters the system may reemerge again in a stable state after some random time.
Could this behavior be related to the evolution of chronic diseases, epileptic seizures, and evolution of the human society?
The authors believe that this question is worthy of investigation. 

In this paper the author considered the notion of the universal fractional $\alpha$-families of maps depending on the non-linearity parameter $K$ and the memory parameter $\alpha > 0$ which is the order of the fractional derivative in the corresponding non-linear fractional differential equation describing a system experiencing periodic kicks. 
$\alpha$-families of maps correspond to a very general form of multi-dimensional non-linear maps with power-law memory in which the present state at time $t$ depends on the previous state at time $t_i$ with a weight proportional to $(t-{t_i})^{\alpha-1}$. 

Two important implementations of this notion are the Standard and Logistic $\alpha$-families of maps. Behavior of these maps is investigated for $\alpha \in [0,2]$. This value of the memory parameter is relevant to the systems with memory which appear in biology and psychology. 
The author shows that considered maps with memory exhibit some universality in the dependence of their properties on $K$ and $\alpha$. 
Phase space of the considered maps, depending on the parameters, may contain periodic sinks, attracting slow diverging trajectories, attracting accelerator mode trajectories, chaotic attractors, and cascade of bifurcations type trajectories whose properties are different from properties of attractors in regular dynamical systems \cite{FM1,FM2,FM3,FM4,FM5,FM6,FM7}.

All natural systems are open systems. An example of an open system with memory, fractional dissipative Standard Map, was considered in \cite{FM2}. The authors of this paper show that open systems with memory demonstrate more complicated properties and more complicated cascade of bifurcations type trajectories than corresponding conservative systems.

\section*{Acknowledgment}
The author acknowledges support from the Joseph Alexander Foundation,
Yeshiva University.  
The author expresses his gratitude to E. Hameiri and H. Weitzner 
for the opportunity to complete this work at the Courant Institute     
and to V. Donnelly for technical help.




\begin{thebibliography}{1}


\bibitem{LRBook} A.~C.~J. Luo and V. Afraimovich (Eds.), 
\textit{Long-range Interaction, Stochasticity and Fractional Dynamics},
New York, Springer, 2010.

\bibitem{LR1} N. Laskin and G.~M. Zaslavsky, ``Nonlinear fractional
dynamics on a lattice with long-range interactions'', \textit{Physica A}
\textbf{368}, 38--4, 2006.

\bibitem{LR2} G.~M. Zaslavsky, M. Edelman, and V.~E. Tarasov, ``Dynamics of the chain of forced oscillators with long-range interaction:
From synchronization to chaos,'' \textit{Chaos}, \textbf{17},
043124, 2007.


\bibitem{ZBook} G.~M. Zaslavsky, \textit{Hamiltonian Chaos and
Fractional Dynamics}, Oxford, Oxford University Press,  2005.


\bibitem {ZEN} G.~M. Zaslavsky, M. Edelman, and B.~A. Niyazov, 
``Self-Similarity,
Renormalization, and Phase Space Nonuniformity of Hamiltonian Chaotic
Dynamics'', \textit{Chaos}, \textbf{7}, 159--181, 1997.


\bibitem{TBook} V.~E. Tarasov, \textit{Fractional Dynamics:
Application of Fractional Calculus to Dynamics of Particles, Fields and
Media}, Beijing, Berlin, HEP, Springer, 2010.


\bibitem {TD1} V.~E. Tarasov,
``Fractional equations of Curie-von Schweidler and Gauss laws'',
\textit{Journal of Physics: Condensed Matter}, \textbf{20}, 145212, 2008. 

\bibitem {TD2} V.~E. Tarasov,
``Universal electromagnetic waves in dielectrics'',
\textit{Journal of Physics: Condensed Matter}, \textbf{20}, 175223, 2008. 

\bibitem {TD3} V.~E. Tarasov,
``Fractional integro-differential equations for electromagnetic 
waves in dielectric media'',
\textit{Theor. and Math. Phys.}, \textbf{158}, 355--359, 2009.


\bibitem{MBook} F. Mainardi, \textit{Fractional Calculus and
Waves in Linear Viscoelasticity: An Introduction to Mathematical Models},
London, Imperial College Press, 2010.

\bibitem {VR1} M. Caputo and F. Mainardi, 
``Linear models of dissipation in anelastic solids'',
\textit{Rivista del Nuovo Cimento}, \textbf{1}, 161--198, 1971.

\bibitem {VR2}  R.~L. Bagley and P.~J. Torvik, 
``A Theoretical Basis
for the Application of Fractional Calculus to Viscoelasticity'',
\textit{Journal of Rheology}, \textbf{27}, 201--210, 1983.

\bibitem {VR3} R.~L. Bagley and P.~J. Torvik, 
``Fractional Calculus - A Different Approach to the Analysis of 
Viscoelastically Damped Structures'', 
\textit{AIAA Journal},  \textbf{21}, 741--748, 1983.

\bibitem {VR4} F. Mainardi and R. Gorenflo,
``Time fractional derivatives in relaxation processes: a tutorial survey'',
\textit{Frac. Calc. Appl. Anal.}, \textbf{10}, 269--308, 2007. 



\bibitem{CB} R. Caponetto, G. Dongola, and L. Fortuna, 
\textit{Fractional Order Systems: Modeling and Control Applications (World
Scientific Series on Nonlinear Science Series a)}, 
London,  World Scientific, 2010.




\bibitem
{BV1} Y. Kobayashi, H. Watanabe, T. Hoshi, K. Kawamura, 
and M.~G. Fujie,    
``Viscoelastic and Nonlinear Liver Modeling for Needle Insertion Simulation'',
\textit{Soft Tissue Biomechanical Modeling for Computer Assisted Surgery,
Studies in Mechanobiology, Tissue Engineering and Biomaterials},
\textbf{11}, 41--67, 2012.
                       
\bibitem {BV2} T.~C. Doehring, A.~D. Freed, E.~O. Carew, and I. Vesely,
``Fractional order viscoelasticity of the aortic valve cusp:
An alternative to quasilinear viscoelasticity'',
\textit{J. Biomech. Eng.}, \textbf{127},  700--708, 2005.


\bibitem{BV3} E. Mace, I. Cohen, G. Montaldo,  and R. Miles, 
``In Vivo Mapping of Brain Elasticity 
in Small Animals Using Shear Wave Imaging'', 
\textit{IEEE Transactions on Medical Imaging}, \textbf{30}, 550--558, 2011.

\bibitem {BV4} S. Nicolle, L. Noguera,  and  J.-F. Paliernea,  
``Shear mechanical properties of the spleen: 
Experiment and analytical modelling'',
\textit{Journal of the Mechanical Behavior of Biomedical Materials},  
\textbf{9}, 130--136, 2012.

\bibitem {BV5} N.~M. Grahovac and M.~M. Zigic, 
``Modelling of the hamstring muscle group by use of fractional derivatives'', 
\textit{Computers and Mathematics with Applications}, \textbf{59},
1695--1700, 2010.


\bibitem {BV6} K. Hoyt, B. Castaneda, M. Zhang, P. Nigwekar,    
A. di Sant’Agnese, J.~V. Joseph, J. Strang, D.~J. Rubens,  and 
K.~J. Parker,  
``Tissue elasticity properties as biomarkers for prostate cancer'',
\textit{Cancer Biomarkers}, \textbf{4}, 213--225, 2008.


\bibitem{Neuron1} H.~Z. Shouval,
``Models of synaptic plasticity'', \textit{Scholarpedia},  \\
http://www.scholarpedia.org/article/Models\_of\_synaptic\_plasticity.


\bibitem{Neuron2} E. Marder, L.~F. Abbott, G.~G. Turrigiano, Z. Liu, and
  J. Golowash,
``Memory from the dynamics of intrinsic membrane currents'',
\textit{Proc. Natl. Acad. Sci. USA},
\textbf{93},  13481--13486, 1996.


\bibitem{Neuron3} B.~N. Lundstrom, A.~L. Fairhall, and M. Maravall, 
``Multiple time scale encoding of slowly varying whisker stimulus
  envelope incortical and thalamic neurons in vivo'', 
\textit{J. Neuroscience}, \textbf{30},
5071--5077, 2010.

\bibitem{Neuron4} B.~N. Lundstrom, M.~H. Higgs, W.~J. Spain, and 
A.~L. Fairhall, 
``Fractional differentiation by neocortical pyramidal neurons'', 
\textit{Nature Neuroscience}, \textbf{11},  1335--1342, 2008.



\bibitem{Magin} R.~L. Magin, 
``Fractional calculus models of complex dynamics in biological tissues'',
\textit{Comp. Math. Appl.}, \textbf{59},  1586--1593, 2010.


\bibitem
{Ada1} A.~L. Fairhall, G.~D. Lewen, W. Bialek,   and
R.~R  de Ruyter van Steveninck,
``Efficiency and Ambiguity in an Adaptive Neural Code'',
\textit{Nature},  \textbf{412}, 787--79, 2001.


\bibitem
{Ada2} D.~A. Leopold, Y. Murayama,   and N.~K. Logothetis, 
``Very slow activity fluctuations in monkey visual cortex: 
implications for functional brain imaging'', \textit{Cerebral Cortex},
\textbf{413}, 422--433, 2003.


\bibitem {Ada3} A. Toib, V. Lyakhov,   and S.   Marom,   
``Interaction between duration of activity and 
recovery from slow inactivationin mammalian brain Na+ channels'', 
\textit{Journal of Neuroscience},  \textbf{18}, 1893--1903, 1998.



\bibitem
{Ada4} N. Ulanovsky, L. Las, D. Farkas,  and I.  Nelken,   
``Multiple time scales of adaptation in auditory cortex neurons'', 
\textit{Journal of Neuroscience}, \textbf{24}, 10440--10453, 2004.

\bibitem
{Ada5} M.~S.  Zilany, I.~C. Bruce, P.~C. Nelson,  
and L.~H. Carney,  
``A phenomenological model of the synapse between the inner hair cell and
auditory nerve: long-term adaptation with power-law dynamics'',
\textit{Journal of the Acoustical Society of America}, \textbf{126},
2390--2412, 2009.

\newpage


\bibitem
{Anderson}  J.~R. Anderson,
\textit{Learning and memory: An integrated approach}, New York, Wiley, 1995.



\bibitem
{Kahana} M.~J. Kahana,  \textit{Foundations of human memory}, 
New York, Oxford University Press, 2012.


\bibitem {F1} D.~C. Rubin and A.~E. Wenzel,   ``One Hundred Years of 
Forgetting: A Quantitative Description of Retention'', 
\textit{Psychological Review}, \textbf{103}, 743--760, 1996.

\bibitem
{F2} J.~T. Wixted,  ``Analyzing the empirical course 
of forgetting'', 
\textit{Journal of Experimental Psychology: Learning, Memory, and
Cognition},  \textbf{16}, 927--935, 1990.

\bibitem
{F3}  J.~T. Wixted and E. Ebbesen,   ``On the form of forgetting'', 
 \textit{Psychological Science},  \textbf{2},  409--415, 1991. 

\bibitem
{F4} J.~T. Wixted and E. Ebbesen,    
``Genuine power curves in forgetting'', 
\textit{ Memory \& Cognition},  \textbf{25}, 731--739, 1997. 




\bibitem{FM1} M. Edelman and V.~E. Tarasov, 
``Fractional standard map'',
\textit{Phys. Let. A}, \textbf{374}, 279--285, 2009.

\bibitem{FM2} V.~E. Tarasov and M. Edelman,
``Fractional dissipative standard map'',
\textit{Chaos}, \textbf{20}, 023127, 2010.

\bibitem{FM3} M. Edelman,
``Fractional Standard Map: Riemann-Liouville vs. Caputo'',
\textit{Commun. Nonlin. Sci. Numer. Simul.}, \textbf{16}, 4573--4580, 2011.

\bibitem {FM4} M. Edelman and L.~A. Taieb, 
``New Types of Solutions of Non-Linear Fractional Differential Equationsin'',
in: \textit{Advances in Harmonic 
Analysis and Operator Theory; Series: Operator Theory: 
Advances and Applications}, Eds.: A. Almeida, L. Castro, and F.-O. Speck,
 \textbf{229}, 139--155, Basel, Springer, 2013.

\bibitem {FM5}  M. Edelman, 
``Fractional Maps and Fractional Attractors. Part I:
        $\alpha$-Families of Maps'',
\textit{Discontinuity, Nonlinearity, and Complexity}, 
\textbf{1}, 305--324, 2013.

\bibitem {FM6} M. Edelman, ``Universal Fractional Map and Cascade of
  Bifurcations Type Attractors'', \textit{Chaos}, \textbf{23}, 033127, 2013. 

\bibitem {FM7} M. Edelman, 
``Fractional Maps as Maps with Power-Law Memory'', 
in: \textit{Nonlinear Dynamics and Complexity}, 
Eds.: A. Afraimovich, A.~C.~J. Luo, and X. Fu, 
79--120, New York, Springer, 2014.




\bibitem{FM8} V.~E. Tarasov and G.~M. Zaslavsky,
``Fractional equations of kicked systems and discrete maps'',
\textit{J. Phys. A}, \textbf{41}, 435101, 2008. 


\bibitem{FM9} V.~E. Tarasov,
``Discrete map with memory from fractional differential 
equation of arbitrary positive order'',
\textit{Journal of Mathematical Physics}, \textbf{50}, 122703, 2009.



\bibitem {Chirikov} B.~V. Chirikov, 
``A universal instability of many dimensional oscillator systems'', 
\textit{Physisc Reports}, \textbf{52}, 263--379, 1979.



\end{thebibliography}
%

\end{document}